\documentclass[manuscript]{acmart}

\usepackage{adjustbox}
\usepackage{array}
\usepackage{multirow}
\usepackage{rotating}
\usepackage{xcolor}
\definecolor{applegreen}{rgb}{0.0, 0.50, 0.0}

\AtBeginDocument{%
  \providecommand\BibTeX{{%
    \normalfont B\kern-0.5em{\scshape i\kern-0.25em b}\kern-0.8em\TeX}}}

\setcopyright{acmcopyright}
\copyrightyear{2023}
\acmYear{2023}
\begin{document}



\title[Are You Worthy of My Trust?]{Are You Worthy of My Trust?: A Socioethical Perspective on the Impacts of Trustworthy AI Systems on the Environment and Human Society}

\author{Jamell Dacon}
\email{jamell.dacon@morgan.edu}
\orcid{0000-0002-6917-6835}
\affiliation{%
  \institution{Morgan State University}
  \city{Baltimore}
  \state{Maryland}
  \country{USA}
}




\begin{abstract}
With ubiquitous exposure of AI systems today, we believe AI development requires crucial considerations to be deemed trustworthy. While the potential of AI systems is bountiful, though, is still unknown--as are their risks. In this work, we offer a brief, high-level overview of societal impacts of AI systems. To do so, we highlight the requirement of multi-disciplinary governance and convergence throughout its lifecycle via critical systemic examinations (e.g., energy consumption), and later discuss induced effects on the environment (i.e., carbon footprint) and its users (i.e., social development). In particular, we consider these impacts from a multi-disciplinary perspective: computer science, sociology, environmental science, and so on to discuss its inter-connected societal risks and inability to simultaneously satisfy aspects of well-being. Therefore, we accentuate the necessity of holistically addressing pressing concerns of AI systems from a socioethical impact assessment perspective to explicate its harmful societal effects to truly enable humanity-centered Trustworthy AI. 

\end{abstract}



\maketitle

\section{Introduction}\label{sec:intro}

Artificial intelligence (AI), a scientific term coined in the 1956 Dartmouth Conference \cite{article2005, article2006}, is a revolutionary science that provides the ability for machines to perform both comprehensible and fundamental functionalities of the human brain. Representative cases of AI applications include but are not limited to: logical reasoning and understanding, learning, problem-solving, agent (human and artificial) interaction, perception and expression of creativity. Within the last decade, modern human society has witnessed the rise of swift and extraordinary developments in AI technology. Today, AI systems are ubiquitous and can simultaneously consider several modalities i.e., visual, auditory and textual; for example, e-commerce platforms' ability to utilize previous search and purchase history to provide personalized recommendations to users based on their online behavior. 

As a result, the widespread integration of AI is becoming increasingly predominant in many real-world applications such as computer vision \cite{voulodimos2018deep}, image generation \cite{van2016conditional, workshop2023bloom}, speech and audio processing \cite{furui2018digital, nassif2019speech}, natural language processing \cite{openai2023gpt4} and recommendations \cite{zhang2019deep}. Thus, generating strong technological waves within interdisciplinary applications ranging from transportation, biometrics, customer service, etc. \cite{ergen2019artificial, BeneficialAI, chowdhery2022palm, FairAI}. Therefore, these facets of AI increase social and economic benefit by having numerous innovative and trustworthy applications in critical domains and other potential aspects for social good. The full scope of these exemplary cases, though, is still unknown---as are their risks. 

Today, large language models (LLMs) and generative AI offer astounding potentialities, constantly expanding the frontier of an algorithmic society \cite{openai2023gpt4, workshop2023bloom, chowdhery2022palm, touvron2023llama}. However, these potentialities are accompanied by existential risks associated with inadvertent overuse or willful misuse of AI technologies which arise largely from contrasting geopolitics and unintended consequences, and relate typically to good intentions gone awry \cite{solaiman2023evaluating, ethicalAI}.
Consequently, recent studies and utilization indicate that AI systems may not be trustworthy as they can potentially produce adverse short and long-term societal effects on people, society and the environment such as unethical use and misaligned incentives---for example, discriminating on the basis of demographic features, malicious intent, or the disposal of dangerous chemicals of AI-related technologies \cite{solaiman2023evaluating, 10.1145/3546872, wu2023unveiling, li2023making, castaño2023exploring, 10.1145/3381831}.
Thus, causing global concern and demand urgent attention.  

\begin{figure}[ht]
    \centering
    \includegraphics[scale=0.4]{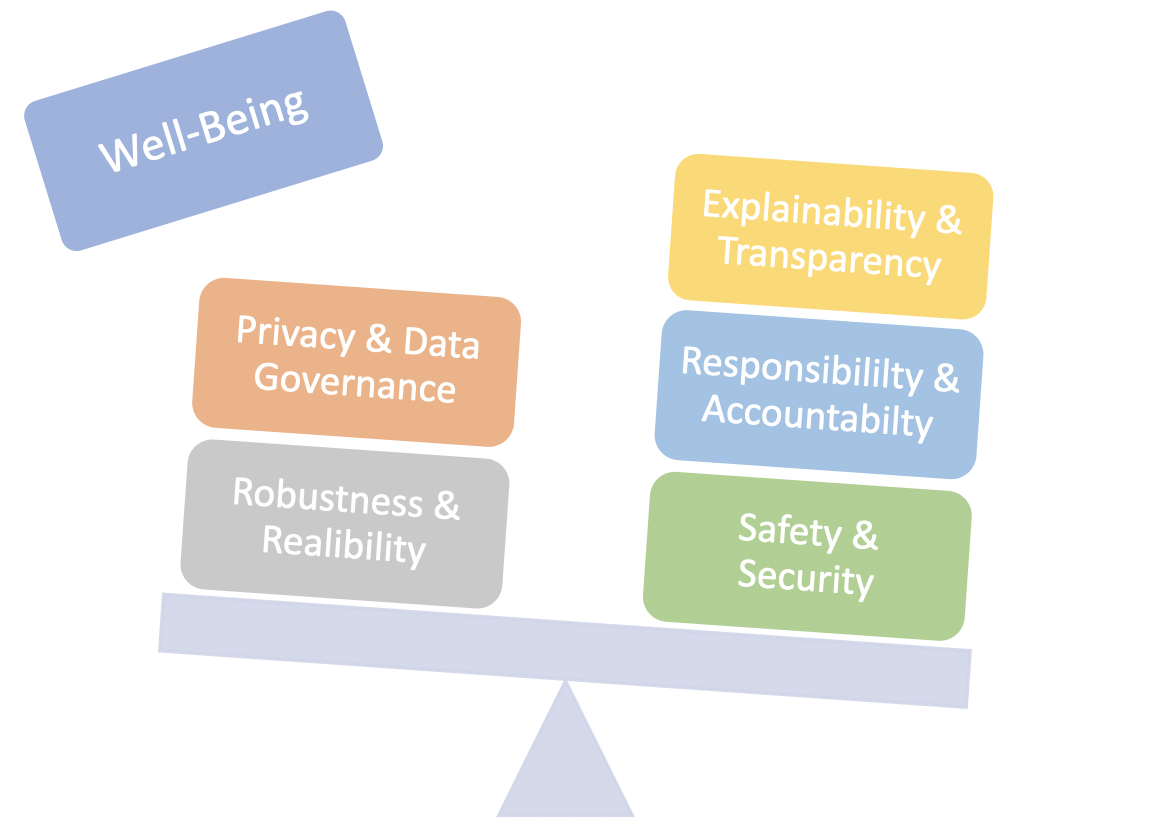}
    \caption{An unbalanced representation of Trustworthy AI given its six key dimensions.}
    \label{fig:untrustworth_ai}
\end{figure} 

According to the Merriam-Webster Dictionary, the first known use of the word ``trustworthy'' (or ``trust'' + ``worthy'') was in 1658. It is esteemed as ``worthy of confidence : dependable''; and defined as ``worthy of trust or confidence; reliable, dependable'' in the Oxford English Dictionary. Over the last decade, this term was coupled with ``artificial intelligence'' to create a new and innovative scientific term to label our trust and dependability on these transformative systems \cite{buccinca2021trust}, which lay an important foundation between human values and AI functionality. Due to a myriad of questions and discussions of what makes AI trustworthy, in this work, we define Trustworthy AI as \textit{humanity-centered application systems that incorporate socioethical factors designed to simultaneously benefit both people and society while posing no risk of potential harm.} According to a book titled, ``Design for a Better World: Meaningful, Sustainable, Humanity Centered'' \cite{norman2023design}, published by the MIT Press 2023 states:
    \begin{quote}\small{Humanity-centered design represents the ultimate challenge for designers to help people improve their lives. Where ``human-centered'' puts a face to the user, ``humanity-centered'' expands this view far beyond: to the societal level of the world population who face hordes of highly complex and interrelated issues that are most often tangled up in large, sophisticated, ``human-centered'' systems.}
    \end{quote}
Overall, Trustworthy AI \textit{must} efficiently encompass a set of factors that are sustainability-focused: to benefit all human beings including future generations and promote globally- and eco-friendliness \cite{10.1145/3381831, Smuha, BeneficialAI}.

As the rapid development and widespread use of AI applications in diverse sectors continue to surge, concerns have emerged about the trustworthiness of AI. Notably, these systems also expose society to potential risks of malicious intentions conjoined with the capabilities of technology, infrastructural development and maintenance, disposal, and greenhouse gas emissions (including carbon dioxide and methane). We posit that it is imperative to highlight both \textit{outcome} (e.g., deliberate and unintentional harms, loss of life, air quality, and climate change) and \textit{error} (e.g., poor system performance and non-interpretable results) disparities of AI systems across six dimensions: (i) \textbf{Safety \& Security}, (ii) \textbf{Robustness \& Reliability}, (iii) \textbf{Responsibility \& Accountability}, (iv) \textbf{Explainability \&  Transparency}, (v) \textbf{Privacy \& Data Governance} and finally (vi) \textbf{Well-Being} as seen in our Trustworthy AI framework.
Despite undeniable benefits of current AI systems disclosing the limitations and risks of innovative AI applications affecting notions of human agency and behavior, globally- and eco-friendliness often assume a secondary position or status in AI research as seen in Figure~\ref{fig:untrustworth_ai}. We believe it is important to disclose this research in full.

In this work, we offer a brief, high-level overview of adverse societal impacts of AI systems, specifically, we consider its \textit{Well-Being} dimension. 
Our goal is to qualitatively accentuate its pitfalls to explicate that AI requires multi-disciplinary governance and convergence, and cross-sectoral cooperation throughout its entire supply chain from ideation to deployment to oversight of human behavioral concepts (e.g., morals), and technical and regulatory compliance. In addition to definitions and concepts, our focal points are to (i) highlight its hidden environmental dilemma (e.g., global warming and harming natural ecosystems), with regard to resource usage, energy and water consumption, and greenhouse gas emissions, and (ii) discuss generated ramifications in psychology and sociology in human society affecting social conceptions of agency, mental health (e.g, emotional, and psychological well-being) and development (e.g., cohesion, progression and welfare) to truly enable humanity-centered Trustworthy AI. This perspective distinguishes this work from existing previous works. In particular, recommendations of technical mitigation techniques and methods are out of scope as we aim to improve understanding (and not take action) of societal impacts of AI. Moreover, we present this work to four groups of readers: (a) in primis users, (b) auditors, (c) legislators and policymakers, and lastly (d) researchers and developers.
Furthermore, we examine each dimension within three contexts: (i) systemic i.e., AI application systems, (ii) individual i.e., user, and (iii) societal i.e., social, and environment.

\section{Related Work} \label{sec:related}

Existing works present trustworthy AI from several overlapping concepts and interpretations such as ethics, sustainability and efficiency \cite{ethicalAI, Smuha, 10.1145/3381831, 17goals}, explainability \cite{ExplainableAI, ergen2019artificial}, fairness and responsibility \cite{FairAI, ResponsibleAI}, and so on. Typically, AI papers tend to target robustness and fairness, particularly with respect to its ability to obtain ``state-of-the-art''\footnote{Meaning, in practice, an explicitly well-formed AI development and robust evaluation process on a downstream task on a new or existing benchmark, achieving similar or greater accuracy than previously reported.} results i.e., judgments, predictions, recommendations, or decisions based on raw data, models or fine-tuning. Therein, computational solutions and qualitative methods (such as tool-kits, repositories and surveys) for supporting end-to-end (i.e., capabilities and vulnerabilities) research in specific dimensions of trustworthy AI are extending and expanding constantly. Many aim to identify technical systemic issues such as
infrastructural disparities, for example, robustness \cite{10.1007/978-3-030-93736-2_33, Xu2020YetML, 10.1145/3447556.3447566, 
pmlr-v139-xu21b}, data leakages and breaches, and unveiling harmful socially-constructed and algorithmic biases \cite{liu-etal-2020-gender,  may-etal-2019-measuring, DBLP:journals/corr/abs-1809-02208}, stereotypes \cite{10.5555/3157382.3157584} and representations harms \cite{10.1145/3442442.3452325} of underrepresented protected classes \cite{dacon2022detecting, tatman17_interspeech, linguistics-011619-030556, race_survey}. Others aim to mitigate harmful impacts such as fairness, security, privacy, and ethical concerns \cite{10.1145/3546872, zou2023universal, 10109301, wu2023unveiling}.

With the increasing demand for disclosing environmental and social effects due to its designated secondary position or status in the AI community, Schwartz and Dodge et al. \cite{10.1145/3381831} introduce the ideology of {\color{red}{RED AI}}---the relationship between model efficiency and the increased unsustainable cost of producing state-of-the-art results. The authors later introduced its solution, namely, {\color{applegreen}{GREEN AI}}---the concept of reducing computational cost of AI research while yielding novel results.
From an environmentally-friendly perspective, Castaño et al. \cite{castaño2023exploring} explores the carbon footprint of machine learning (ML) models, whereas Li et al. \cite{li2023making} focuses on uncovering and addressing the water footprint of AI models.
In a recent work, Solaiman and Talat et al. \cite{solaiman2023evaluating} focus on evaluating the social impact of generative AI systems in base technical systems, and in people and society. The authors specify seven categories of social impact from a systemic perspective: bias, stereotypes, and representational harms; cultural values and sensitive content; disparate performance; privacy and data protection; financial costs; environmental costs; and data and content
moderation labor costs. Additionally, they present a societal perspective offering five overarching categories of what can be evaluated in society: trustworthiness and autonomy; inequality, marginalization, and violence; concentration of
authority; labor and creativity; and ecosystem and environment. 

In tandem with the increase in literature on evaluating AI-related socially-constructed biases, risks and impacts \cite{ solaiman2023evaluating, hovy-spruit-2016-social, dacon2022detecting}
sociotechnical harms \cite{shelby2023sociotechnical, king2020artificial} in areas of natural language processing (NLP) and generative AI, the 
societal impact problem considered by our paper, particularly with regard to highlighting AI systems' hidden environmental dilemma (i.e., excessive resource usage and deterioration of natural ecosystems) and discussing generated ramifications in the mental health and sociology of its consumers are significant and have not been adequately addressed within or by the AI community. As adverse societal impacts caused by AI may not be fully quantifiable due to its complexity, we adopt a proactive approach from several disciplines such as computer science, social science, psychology, environmental science, education, health, chemical and resource engineering, and law to better grasp the vulnerabilities of such systems. Moreover, we accentuate the necessity of cohesively addressing pressing concerns of AI systems from a multi-disciplinary socioethical perspective to minimize the foundation gap between human values and AI functionality, and thus, truly enable humanity-centered Trustworthy AI. 

\section{Background} \label{sec:concepts}

To build a Trustworthy AI system, it is crucial to address concerns around the widespread use and integration of AI application systems throughout its lifecycle from ideation to design, development, deployment to oversight of unintended consequences (i.e., impacts and risks) and regulatory compliance. For example, a common  misuse case is \textit{information disorder} (e.g., misinformation, disinformation and malinformation) which can lead to mass production of social and mediatic discourse. Therefore, the aim of Trustworthy AI is to encompass a unified framework of fundamental principles to display its robustness, reliability and explainability while being fair, safe and secure, to ensure beneficial impacts on both human society (including future generations) and the environment.

As Trustworthy AI continues to garner massive attention and can be interpreted from several overlapping concepts, including Green AI \cite{10.1145/3381831}, Ethical AI \cite{ethicalAI}, Explainable AI \cite{ExplainableAI}, Fair AI \cite{FairAI}, Responsible AI \cite{ResponsibleAI} and so on, we provide concepts and definitions to articulate its fundamental principles. Note that the six aforementioned key dimensions in this framework are \textit{not} mutually exclusive. To help readers (in primis users, developers and legislators) understand and verify applicable components of each dimension, we contribute a comprehensive list of systemic and societal requirements that AI systems should meet in order to be deemed trustworthy in Table~\ref{tab:factors}, categorized by their root principles. These principles are (i) \textbf{Harm Prevention}, (ii) \textbf{Explicability} and (iii) \textbf{Fairness}.

\subsection{Harm Prevention}
\subsubsection{Safety \& Security}\label{sec:safesec}  
An integral component of achieving Trustworthy AI is its safety and security protocols for potential physical and cyber risks. A trustworthy AI system should have safeguards that enable a contingency plan that is adaptable, for example, possessing a human-in-the-loop (HITL), human-on-the-loop (HOTL), or human-in-command (HIC) paradigm or capable of switching from a statistical to principle-driven self-alignment rule-based procedure. This necessitates processes to clarify and explain potential risks associated with the systems in downstream tasks. Consequently, this ensures performance as intended without harming humans or the environment. However, the level of safety measures required depends on the magnitude of the potential risks posed by an AI system, which in turn depends on the system's capabilities.
Powerful AI systems may pose higher risks, thus, it is crucial for safety measures to be developed and tested proactively. Moreover, a trustworthy AI system should consist of general safety compliance and protocols to secure against unintentional and unexpected risks preventing unacceptable harm, for example, adversarial exploitation against vulnerabilities, namely, cyberattacks.

\begin{table}[ht]
    \caption{A comprehensive list of contributing factors in our Trustworthy AI framework categorized by their overarching principles.}\label{tab:factors}
\scalebox{0.98}{\begin{tabular}{|c|cc|c|}
\hline
\textbf{Principles} & \multicolumn{2}{c|}{\textbf{Dimensions}} & \textbf{Properties} \\ \hline
\multirow{3}{*}{\textit{Harm Prevention}} & \multicolumn{2}{c|}{Safety \& Security} & \begin{tabular}[c]{@{}c@{}}User Friendly, Accessible, Equitable, \\ Inclusive, Unbiased, User Protection, Invulnerable\end{tabular} \\ \cline{2-4} 
 & \multicolumn{2}{c|}{Robustness \& Reliability} & Adaptable, Accurate, Consistent, Predictable \\ \cline{2-4} 
 & \multicolumn{2}{c|}{Privacy} & \begin{tabular}[c]{@{}c@{}}Autonomous, Confidential, Consensual, \\ Discretional, Data Governance, Access \& Protection\end{tabular} \\ \hline
 \textit{Explicability} & \multicolumn{2}{c|}{Explainability \& Transparency} & \begin{tabular}[c]{@{}c@{}}Interpretable, Visible, Justifiable,\\  Ethical, Traceable, Clarifiable, Communicable\end{tabular} \\ \hline 
\multirow{3}{*}{\textit{Fairness}} & \multicolumn{2}{c|}{Responsibility \& Accountability} & \begin{tabular}[c]{@{}c@{}}Auditable, Common \& Social Good, Ownership, \\ Resolvable, Sustainability Focused, Value \& Moral Adding\end{tabular} \\ \cline{2-4} 
 & \multicolumn{1}{c|}{\multirow{2}{*}{Well-Being}} & Social & \begin{tabular}[c]{@{}c@{}}Diverse, Non-subversion, Non-discrimination, \\Equitable, Social \& Cultural Agentive, Social Justice\end{tabular} \\  \cline{3-4} 
 & \multicolumn{1}{c|}{} & \begin{tabular}[c]{@{}c@{}} Environmental  \end{tabular} & Sustainable, Energy-efficient, Viable, Global- \& Eco-Friendliness\\ \hline
\end{tabular}}
\end{table}

\subsubsection{Robustness \& Reliability}\label{sec:robustrel}
For AI systems to be considered trustworthy, they should possess technical robustness and resilience to produce consistent and reliable outputs. This ensures that the potential for malicious and unintentional harm can be minimized and prevented. A reliable AI system must exhibit accuracy in a broad range of situations. Accuracy pertains to an AI system’s ability to make correct judgments, for example to correctly classify information into the proper categories, or its ability to make correct predictions, recommendations, or decisions based on data or models. An explicit and well-formed development and robust evaluation process can support, mitigate and correct unintended risks from inaccurate predictions, particularly with regard to its robustness against perturbations and distribution shifts. When occasional inaccurate predictions cannot be avoided, it is important that the system can indicate how likely these errors are. A high level of accuracy is especially crucial in situations where the AI system directly affects human lives.

\subsubsection{Privacy}

Most AI algorithms, applications and systems are data driven and may require leverage of confidential, discretional and sensitive consumer data (e.g., sexual and gender orientation). As data is tightly associated and paramount to the functionality of these systems, the quality and integrity of data usage, and respect for privacy and data protection is necessary. 
To prevent unlawful acts of malicious human behavior and attacks adequate data governance and safety measures must be \textit{guaranteed}. Specifically, data governance and safety measures of a trustworthy AI system will prevent users from digital harm such as data poisoning, model leakage or underlying infrastructure--both software and hardware, and protect confidential and sensitive user information. 
For users to holistically trust an AI system, user should be provided explicability of how their data is collected, used and extensively contribute to the decision-making process. In addition, throughout the lifecycle of an AI systems, users should be able to opt in and out of sharing their data as many applications of the AI system may be multi-use applications, hence, for an AI systems to be considered secure, cross-platforms data governance, access and protection should be taken into account from ideation to deployment.

\subsection{Explicability}
\subsubsection{Explainability \& Transparency}\label{sec:extrans}
To summarize the reasons of the AI system's decision-making process, gain the trust of its users, and produce insights, trustworthy AI must demand and maintain suitable explanations of its entire supply chain. Typically, in AI literature, interpretability and explainability have been used interchangeably. Explainability differs as explainable models are interpretable by default, but the reverse is not always true \cite{Gilpin2018ExplainingEA}. Specifically, explainability refers to the explanatory ability of both the technical processes of an AI system such as transparency, traceability and clarity of the capabilities and limitations of systemic attributes, and correlations between data collection and human behavior to be disclosed and minimized. A fundamental right of the user is transparency of high-stake decision-making, models and data usage to understand the reasoning behind an AI system's mechanisms. 
Moreover, explicability in AI is crucial as it refers to the detection and explanation of anomalies and model misspecifications.

\subsection{Fairness}
\subsubsection{Responsibility \& Accountability}\label{sec:resacc}

To ensure responsibility and accountability in AI systems, mechanisms such as organizational protocols i.e., audits, reports and policies should be in place to assess algorithms, data and design processes and determine who is responsible for the output of AI system decisions and their outcomes. A key role therein is auditability--identification, assessment and robust evaluation by internal, external and independent auditors to assess in-house systemic infrastructure and their potential outcomes throughout its lifecycle via a critical examination of the resource usage, and energy and water consumption and so on, during its entire supply chain; therefore having some overlap with the principles of explicability. These assessments aid in minimizing the potential harmful impacts or risks that the AI system may pose. In particular, responsibility refers to developers duty to ensure that AI system comply with its legal, ethical and sustainability focused obligations in task-oriented situations; whereas accountability refers to acknowledgment, response and ownership of the developer to redress and resolve adverse impacts to ensure the trustworthiness of the AI system.

\subsubsection{Well-Being}\label{sec:wellbeing}

Well-being can be described as an inter-dimensional component of Trustworthy AI comprised of all 3 principles to avoid unfavorable disparities in human-computer interactions to ensure fairness in the decision-making process for humans and ensure sustainability-focused practices by considering potential societal impacts of AI adoption. Specifically, this dimension contains a conjoint set of societal components i.e., social and environmental factors such as non-discrimination and fairness, environmental sustainability and friendliness, and so on. In particular, ubiquitous exposure, implementation and usage of AI systems calls for the consideration of sociological aspects of social and cultural agency to account for its effects not only on users, but on judicial and political processes as well. In addition, a trustworthy AI system will also include inter-dimensional factors and provide its practitioners and users with a sense of comfort and congeniality, for example, fairness and equity in decision making under uncertainty and limited information.
Below we will detail the definitions of the words \textit{social} and \textit{environmental} to provide readers with contexts:   

\begin{itemize}
    \item \textit{\textbf{Social}.} The word ``social'' is noted to mean ``of or relating to human society, the interaction of the individual and the group, or the welfare of human beings as members of society'' in the Merriam-Webster Dictionary; and characterized as ``of or relating to society, as social background, social climate, social duty, social fabric, social issue, social question, social virtue, etc.'' in the Oxford English Dictionary. 

    \item \textit{\textbf{Environmental}.} In the Merriam-Webster Dictionary, the word ``environmental'' possesses a two-part definition:
    \begin{enumerate}
        \item ``the complex of physical, chemical, and biotic factors (such as climate, soil, and living things) that act upon an organism or an ecological community and ultimately determine its form and survival''; and 
        \item ``the aggregate of social and cultural conditions that influence the life of an individual or community''
    \end{enumerate}
\end{itemize}

\section{Environmental Impacts} \label{sec:envimpacts}

As previously mentioned in Table \ref{tab:factors}, environmental properties of Trustworthy AI are sustainability, viability, energy-efficiency, and global- and eco-friendliness. However, the swift development, widespread adoption, and disposal of AI technology are all accompanied by energy-intensive processes and significant greenhouse gas emissions (i.e., carbon footprint). For example, according to Euro News\footnote{\url{https://www.euronews.com/}} researchers estimate that the lifecycle of GPT-3, a powerful and widely distributed 175 billion-parameter state-of-the-art generative AI system, consumed 1,287 megawatt hours (MWh) of electricity and generated 552 metric tons of carbon dioxide ($\mathrm{CO_{2}}$) which is the carbon dioxide equivalent ($\mathrm{CO_{2}e}$) of 123 gasoline-powered cars constantly being driven for one year \cite{gpt3emission}. 
\begin{table}[ht]
    \caption{A comparative analysis of carbon dioxide equivalent ($\mathbf{CO_{2}e}$) of daily life activities and recent (< 5 years) widely deployed popular open-source and publicly available AI systems trained once from developers Google, Meta and OpenAI, and energy consumption. The statistics are reported in terms of paper release date, processing information such processor type, estimated energy consumption kilowatt hours (KWh), processor hours and carbon emissions (lbs). }\label{tab:consumption}
\begin{tabular}{|c|c|c|c|c|c|}
\hline
 & \textbf{\begin{tabular}[c]{@{}c@{}}Paper Release\\  Date\end{tabular}} & \textbf{\begin{tabular}[c]{@{}c@{}}Energy Consumption\\  (kWh)\end{tabular}} & \textbf{Hours} & \textbf{Type} & $\mathbf{CO_{2}e (lbs)}$\\ \hline
\begin{tabular}[c]{@{}c@{}}Flight (NYC $\leftrightarrow$ PEK) \\ (1 passenger) \end{tabular} & - & - & - & - & 5,009 \\ \hline
\begin{tabular}[c]{@{}c@{}}Human life \\ (avg. 1 year)\end{tabular} & - & - & - & - & 11,023 \\ \hline
\begin{tabular}[c]{@{}c@{}} Car \\ (avg. 1 lifetime)\end{tabular} & - & - & - & - & 126,000 \\ \hline
BERT \cite{DBLP:journals/corr/abs-1810-04805} & Oct, 2018 & 1,507 & 6,400 & TPU & 1,438 \\ \hline
T5 \cite{DBLP:journals/corr/abs-1910-10683} & Oct, 2019 & 86,000 & 245,760 & TPU & 103,618 \\ \hline
GPT-3 \cite{DBLP:journals/corr/abs-2005-14165} & Jun, 2020 & 1,287,000 & 3,552,000 & GPU & 1,216,952 \\ \hline
PaLM \cite{chowdhery2022palm} & Apr, 2022 & 3,436,000 & 8,404,992 & TPU & 597,453 \\ \hline
BLOOM \cite{workshop2023bloom} & Nov, 2022 & 520,000 & 1,082,990 & GPU & 66,139 \\ \hline
LLaMA-2 \cite{touvron2023llama} & Jun, 2023 & 1,400,000 & 3,311,616 & GPU & 1,307,342 \\ \hline
\end{tabular}
\end{table}
In practice, models are typically trained several times during research and development. Therefore, the numbers shown in Table \ref{tab:consumption} are only reflective of these systems being trained once, and does not include post-deployment data center cooling energy and carbon cost of maintenance. 

Consequently, as industrial and research AI systems are accuracy-focused rather that being human- or humanity-centered, they can comprise of billion-parameter state-of-the-art neural architectures which are computationally expensive and energy intensive. Specifically, to outperform existing achievements and minimize error disparities, systems are trained on multiple high-performance tensor processing units (TPUs) and graphics processing units (GPUs), ranging from hours to months of training time directly relating long-term AI environmental effects. Fossil fuel combustion used to power data centers and machine learning operations (MLOps) cause detrimental environmental impacts ranging from inequity, quality of life, accelerating global warming and aggravating other climate changes \cite{different_futures}. We highlight the following high-level, non-exhaustive overarching categories of adversarial environmental impacts: (i) Carbon Footprint, (ii) Energy Consumption \& Water Footprint, and (iii) Impact on Natural Ecosystems.

\subsection{Carbon Footprint}\label{sec:carbonfootprint}

In recent years, a large body of reports and literature on rising average global temperature \cite{17goals, castaño2023exploring, 10.1145/3381831} has emerged. Schwartz and Dodge et al. \cite{10.1145/3381831} describes the expensive and environmentally unfriendly participation in NLP research as a barrier, namely, {\color{red}{RED AI}}, due to its designated secondary position or status attributable to computational inefficiency.
However, there is little known about the absolute $\mathrm{CO_{2}}$ emissions of most widely deployed AI technologies on respective developer sites, platforms or repositories. Recently, Hugging Face Hub \cite{huggingface}, a reproducibility-focused repository started $\mathrm{CO_{2}}$ reporting in 2021.
Subsequently, according to a repository mining study \cite{castaño2023exploring}, this repository currently lacks explainability and transparency of $\mathrm{CO_{2}}$ reporting. 
These reports are essential socioethical considerations when building Trustworthy AI systems adhering to its principles of harm prevention, explicability and fairness. Due to AI integration, there is an increase in $\mathrm{CO_{2}}$ atmospheric concentration, therefore, it is important to recognize that Methane ($\mathrm{CH_{4}}$) is also responsible for a large portion of recent warming despite having a lower global warming potential (GWP) than $\mathrm{CO_{2}}$.

As widespread adoption, and rapid development and deployment of large-scale AI systems (e.g., BERT, GPT-4, LLaMA-2) proliferate, greenhouse gas emissions (including Carbon Dioxide, Methane and Nitrous Oxide) are causing global concerns for human society and the environment. In Table \ref{tab:consumption}, we conduct a comparative analysis of $\mathrm{CO_{2}e}$ (in pounds) of daily life activities and recent widely deployed AI systems as $\mathrm{CO_{2}}$ is made by both natural and man-made processes. Due to the release date of AI systems, we also present their underlying characteristics that also result in adverse environmental impacts. For instance, several informal reports estimate that training GPT-4 (ChatGPT's successor) produces the same $\mathrm{CO_{2}e}$ as 284 gas-powered cars (15,000 metric tons or 35,803,072 pounds) over their lifetime or the equivalent of approximately 7,148 round-trip flights from New York City (NYC) to Beijing (PEK). 

Moreover, these numbers are only conclusive of training emissions and not of updating and supervised fine-tuning (SFT) \cite{sun2023principledriven}, reward modeling and reinforcement learning from human feedback 
(RLHF) \cite{RLHF}, 
and daily consumer query costs and emissions. Therefore, the proliferation of powerful, energy-intensive AI systems significantly contribute to global warming, poor air quality, and massive resource usage---directly aggravating climate change \cite{10.1145/3381831, castaño2023exploring, different_futures}.

\subsection{Energy Consumption \& Water Footprint}\label{sec:water}

Consumption of resources continues to receive momentous attention from academia, industry and government agencies in hopes of accelerating the shift towards a more AI-based society. Typically, data centers generate massive amounts of electricity from fossil fuels combustion which releases greenhouse gases into the atmosphere, leading to an increase in global warming and other climate changes \cite{different_futures}. Therefore, electricity usage is time- and location-agnostic, and thus, directly correlated with carbon emissions further amplifying global warming \cite{10.1145/3381831}. To measure the energy consumption, AI hardware also needs significant consideration in energy consumption \cite{10.1145/3351095.3372873, zhuang2021randomness}. For example, developers train on powerful energy-intensive hardware, namely TPUs (e.g., TPUv4) and GPUs (e.g., A100), which require massive amounts of electricity typically ranging from 1000 to 4000 MWh per month. Moreover, as developing new and diverse model architectures advance rapidly, larger models will perpetuate market. For instance, NDIVIA released an H100 GPU in March 2022, which is 4 times faster than A100 and plan to release H100 TensorRT-LLM which returns an 8 times total increase to deliver the highest throughput compared to A100 \cite{H100TensorRT-LLM}.
The energy consumption of these large-scale AI systems is later compounded by the energy required to keep them running, then exacerbated by constant updates and fine-tuning to address AI dimensional concerns such as privacy, safety \& security, social well-being and so on. 

Consequently, \textit{adaptive design} and \textit{model compression} have been proposed in several works \cite{cheng2020survey, DBLP:journals/corr/abs-1808-01550, DBLP:journals/corr/abs-1909-11942} to reduce the size of an AI model by considering model architecture, storage space and energy consumption during training and deployment. However, this is not an end-all be-all solution. For example, both GPT-3 and BLOOM \cite{workshop2023bloom} have a model size of $\sim$175B parameters, trained on 300-350B tokens on NVIDIA GPUs, yet BLOOM produces a carbon grid intensity (kg$\mathrm{CO_{2}e}$/ KWh)\footnote{Kilograms to Pounds. 1 kilogram (kg) is equal to 2.205 pounds (lbs).} of 0.057 while GPT-3 produces a carbon grid intensity of 0.429. Although, BLOOM is trained on a larger dataset, it consumes
60\% less energy and emits 18.4 times less $\mathrm{CO_2}e$ than GPT-3. Specifically, BLOOM took 70\% less processor hours due to utilizing a newer (and more efficient) A100 GPU compared to an older V100 GPU used in the training process of GPT-3. Therefore, \textit{hardware} also requires deeper consideration as energy-efficient factor in building Trustworthy AI, evoking the questions surrounding the topics of recyclability and environmental degradation.  

Furthermore, the \textit{water footprint} i.e., water used to cool AI severs is an under-examined topic in diverse AI sectors, thus, lacking transparency, responsibility and accountability. A recent study titled, ``Making AI Less "Thirsty"'' \cite{li2023making}, a first-of-its-kind study, comprehensively highlights the water footprint claiming:
\begin{quote}
    ``\textit{training GPT-3 in Microsoft's state-of-the-art U.S. data centers can directly consume 700,000 liters (185,000 gallons) of clean freshwater (\textbf{enough for producing 370 BMW cars or 320 Tesla electric vehicles})\footnote{Code is available at: \url{https://github.com/RenResearch/Making-AI-Less-Thirsty} to replicate the findings.} and the water consumption would have been tripled if training were done in Microsoft's Asian data centers, according to public data sources \cite{bmwgroup, teslagroup, DBLP:journals/corr/abs-2104-10350, walsh} ... and GPT-3 is also responsible for an additional off-site water footprint of 2.8 million liters due to electricity usage}'' 
\end{quote}
Nevertheless, water-usage is highly dependent on several factors, including data center location and design, water cooling system, and regional climate (e.g., U.S. vs. France). As freshwater scarcity continues to be one of the most pressing global challenges, particularly with regard to a rapidly growing population, as well as the increasing complexity of datasets and models. AI developers must be socially and environmentally responsible and accountable for and transparent about their water footprint to truly enable Trustworthy AI.

\subsection{Impact on Natural Ecosystems}\label{sec:naturaleco}

The past decade has witnessed the rise and huge demand of autonomous and electric vehicles (EVs), normalizing the convenience of driver-less automobiles and providing swift delivery of goods and services. Although, EVs produce less carbon emissions than gas-powered vehicles, there has been an accelerated growth of the lithium ion battery (LIB) market raising ethical questions about their end of life and recyclability \cite{lithiumbatteries}. Recently, direct recycling techniques have been proposed and developed through various pyrometallurgical and hydrometallurgical mechanical, chemical and physical processes causing detrimental environmental impacts. According to \cite{lithiumbatteries}, prior to their work, there were no state-of-the-art direct recycling process studies of closed-loop (i.e., cradle-to-grave) life cycle assessment (LCA) of EV lithium-ion batteries, nor discussions of the saving potential of raw materials which can be used in manufacturing of new high-quality batteries for EVs. 

\noindent{\textit{\textbf{Electronic Waste \& Disposal.}}}
Additionally, energy waste also known as ``E-waste'' produced by AI technology is an under-examined area in AI research conducted by the computer science community. Specifically, E-waste contains hazardous chemicals which include but not limited to: lead, mercury, and cadmium. Contamination of soil and water supplies by these chemicals can have detrimental impacts on natural ecosystems i.e., endangering both human health and the environment. Moreover, the environmental harm caused by the release or leakage of dangerous compounds of AI-related technologies has not yet drawn sufficient attention which is not acceptable. Therefore, to reduce its negative environmental effects and ensure global- and eco-friendliness, critical examinations, laws and socioethical disposal practices concerning e-waste management and proper recycling of AI-related technologies requires considerable attention.

\noindent{\textit{\textbf{Lack of Foresight.}}} As humanity advances towards a more algorithmic society, the utilization of constant innovative technologies usually occurs with little consideration of long-term disadvantages. The global exposure, implementation and utilization of lithium ion batteries (LIBs) is an excellent example of this \cite{kim2019lithium}. The first commercial LIB was issued in 1991, since then, LIBs have monopolized the majority of the electronic market \cite{kim2019lithium, li201830}. Nevertheless, when first released, its unethical mining practices and unsteady recycling options were often disregarded. As AI technology continues to advance rapidly, the demand for LIBs increase. Moreover, the excessive material collection, part production, and the environmental impact of lithium and cobalt mining still have not garnered similar attention \cite{li2011materials, asadi2021review}. As a result, there has been a lack of foresight into its viability, and global and eco-friendliness due to a primary focus as a promising energy storage technology.

\section{Social Impacts} \label{sec:socialimpacts}

The goal of AI should be to foster human nature and create, not obscure human intelligence nor disrupt social conceptions of agency and development such as social cohesion. Yet, current debates question its adversarial social impact on human society. For instance, questions such as: ``\textit{By whom}, \textit{how}, \textit{where}, and \textit{when} this positive or negative impact will be felt?'' receives continuous attention from both academia and industry \cite{ethicalAI}. We will now present different definitions of \textit{social impact} from three disciplines: (i) \textbf{Psychology}; (ii) \textbf{Sociology}; and (iii) \textbf{Computer Science} (technical).
\begin{enumerate}
    \item \textit{Psychology}: The \textit{social impact theory} in psychology refers to the amount of influence a person experiences in group settings that are determined by (a) social status, (b) immediacy, and (c) the number of sources \cite{latane1981psychology, latane1981social, sedikides1990social}. 

    \item \textit{Sociology}: Becker et al. \cite{becker2001social} defines \textit{social impact} as the process of identifying the future consequences on people and communities of current or proposed action or inaction, activity, project, programme or policy \cite{socimpact}, which are related to individuals, organizations, and social macrosystems.

    \item \textit{Computer Science}: Solaiman and Talat et al. \cite{solaiman2023evaluating} define \textit{social impact} as ``\textit{the effect of a system on people and communities along any timeline with a focus on marginalization, and active harm that can be evaluated}''. 
\end{enumerate}
In each case, AI can be used to create opportunities; however, its unethical use (i.e., inadvertent overuse or willful misuse) and misaligned incentives constitute 
expedient various new and creative AI-related risks \cite{king2020artificial}. This includes but is not limited to physical and digital harm to users (e.g., red-teaming), AI-related accidents (e.g., loss of life) or more abstract issues such as biases, stereotypes, representational harms, etc. These risks can potentially produce adverse short and long-term societal effects on people, social groups, and communities.

Consequently, many legal jurisdictions and regulatory committees propose AI regulation plans and guidelines due to malicious manipulation. For instance, the Federal Trade Commission (FTC) recently opened a civil\footnote{Federal Trade Commission (``FTC'') Civil Investigation File No. 232-3044} investigation \cite{civil} against ChatGPT in July 2023 over potential consumer harm. In particular, ChatGPT is an in-context learning interactive text generating platform released in November 2022. Within 6 months of being released, the platform amassed over 100 million user and over 1.6 billion visits in June 2023 \cite{wu2023unveiling}, hence it is coined ``the fastest-growing consumer app in history'' to date. Subsequently, in an article titled ``Untangling Disinformation'' published in August 2023 by National Public Radio (NPR) mentions that online AI chat-bots from leading AI companies including Google, Meta, and OpenAI contain vulnerabilities which disclose user privacy, and leak confidential and sensitive information by implementing an elementary cybersecurity tactic called ``red teaming''. This technique involves hacking with words in lieu of code and hardware, posing severe \textit{safety}, \textit{security} and \textit{privacy} issues \cite{NPR, zou2023universal}.
 
Similarly to the fundamental points raised in \cite{ethicalAI, hagerty2019global} and categories introduced in \cite{solaiman2023evaluating}, we briefly, extend or expand on these context-specific categories by presenting the following high-level, non-exhaustive overarching categories of adverse social effects (and disadvantages) of AI affecting human development in various sectors and industries as they relate to base principles of Trustworthy AI:
\begin{itemize}
    \item Devaluation and Desensitization 
    \begin{itemize}
        \item[--] Eroding Human Self-Determination and Relevance
        \item[--] Despondency of Human Skills  
    \end{itemize}
    \item Human Inactivity and Removal of Responsibility
    \begin{itemize}
        \item[--] Lack of Authoritative Control and Regulation
        \item[--] Ethical, Moral and Value Misalignment
        \item[--] Discontentment and Overreliance
    \end{itemize}
    \item Social Cohesion and Incoherency 
    \begin{itemize}
        \item[--] Job Market Shift and Displacement 
        \item[--] Disjointedness of Human Control and Privacy
        \item[--] Community Erasure 
    \end{itemize}
    \item Proliferation of Malicious Intent and Manipulation 
  \begin{itemize}
        \item[--] Ethical Concerns
        \item[--] Physical and Digital Harm e.g., Security and Privacy threats
        \item[--] Information Disorder i.e., Misinformation, Disinformation and Malinformation
    \end{itemize}
\end{itemize} 
The above classifications can have overlap in their target (individual vs. social group vs. community), yet constitutes different levels of mental distress. Not that social impacts is not solely a ``data problem''; therefore, it may not be an ``easy or simple'' fix.

\subsection{Devaluation and Desensitization}
\noindent{\textit{\textbf{Eroding Human Self-Determination and Relevance}}}: LLMs and Generate AI systems have garnered massive attention due to their impressive cross- and multi-modal generating capabilities content, Currently, there are out- and in-class debates on the topic of demotivation \cite{overreliance}, for example, the involvement of an autonomous system in judicial decision-making. With AI automating large-scale intricate, tedious and monotonous tasks such as writing essays, emails, legal documents, solving mathematical equations, translating languages, etc., there is a pressing long-term concerns of  unfounded trust \cite{skjuve2021my} and other social disparities e.g., literacy, early childhood education, and community, safety, and social contexts \cite{rieger2022human, hovy-spruit-2016-social}.

\noindent{\textit{\textbf{Despondency of Human Skills}}} :Most AI algorithms, applications and systems are data driven, and thus, are designed to be statistical, and rule-based. By default, AI systems lack creativity, and may not possess an HITL, HOTL, or HIC paradigm resulting in redundancy, and potential lack of originality and authenticity in AI-generated content. However, as the intention of AI is to benefit humans by reducing human labor, and thus, increasing everyday conveniences could potentially cause a workforce shift dictating the future of employment \cite{hovy-spruit-2016-social}, and proliferate human dependency of AI systems \cite{overreliance}.

\subsection{Human Inactivity and Removal of Responsibility}
\noindent{\textit{\textbf{Lack of Authoritative Control and Regulation}}}: With the inadvertent overuse or willful misuse of AI technologies, regulatory bodies have consistently released regulation guidelines skew heavily toward consumer harm \cite{NPR}. However, big tech companies are constantly pushing towards the ideologies of monopolization of consumer data, and domination (or ``power concentration'') of consumer use. This can result in economic inequity, power imbalances, wealth disparities and systemic injustices \cite{overreliance}. A lack of proper authoritative control and regulation guidelines to enforce technical and regulatory compliance can lead to the oversight of creating and promoting of harmful biases, prejudice, discrimination, stereotypes, representations harms, etc., in data collection, biased algorithms and people \cite{10.5555/3157382.3157584, 10.1145/3442442.3452325}. \\

\noindent{\textit{\textbf{Ethical, Moral and Value Misalignment}}}: With the rapid adaptation and progress of AI in a multitude of sectors and industries, there has been pressing concerns and challenges surrounding the implementation of ethics, morality and values in highly autonomous AI systems \cite{rieger2022human, overreliance}. These features are paramount to humanity, directly contributing to a flourishing society and can be challenging to incorporate into an AI-related technology. \\
{\color{red}{\textbf{CASE STUDY}}}: In transportation, liability in the event of loss of life or collision, particularly with regard to autonomous self-driving vehicles has drawn considerable attention recently. Due to failure of fundamental principles explicability and fairness (i.e., responsibility and accountability) of AI-controlled vehicles is a complex and ongoing challenge. \\

\noindent{\textit{\textbf{Discontentment, Dependency and Overreliance}}}: The coupled relationship between mimicry capacities, technical robustness, consistency and its ability to produce reliable outputs increases reliability, hence, gaining the trust of its users. However, excessive trust in AI-related technologies leads to overreliance and dependency \cite{overreliance, buccinca2021trust}. This can negative impact consumers as it can bear a significantly high psychological, sociological and physical cost in a broad range of situations \cite{rieger2022human}, example malicious intent and manipulation
\cite{wu2023unveiling, zou2023universal, pmlr-v139-xu21b}.\\

\subsection{Social Cohesion and Incoherency}
\noindent{\textit{\textbf{Job Market Shift and Displacement}}}: In recent years, there has been a surge in AI automation leading to a substantial number job losses and displacements. According to Tech Crunch, a start up that specializes in technology news reported a running total of 224,503 tech layoffs\footnote{\url{https://techcrunch.com/2023/09/07/tech-industry-layoffs-2023/}} from January 2023 to August 2023 \cite{future}. This number is based on full months in certain sectors including but not limited to: Marketing, Security, Finance, Human Resources, and  Media. \\
{\color{red}{\textbf{CASE STUDY}}}: As section \ref{sec:naturaleco} discusses the environmental impact of lithium and cobalt mining, which is native to the Democratic Republic of Congo---the world's largest producer of cobalt. According to The American Broadcasting Company (ABC News), an American commercial broadcast television network states that, ``\textit{...a measure has been introduced in the U.S. House to ban imported products containing minerals critical to electric vehicle batteries but mined through trafficked workers and child labor and other abusive conditions in Congo}''\cite{congo}. Therefore, through the lens of race and ethnicity, job market shift and displacement constitute to AI-related socioeconomic disparities across a broad range of dimensions (i.e., social class, age, geography).\\

\noindent{\textit{\textbf{Disjointedness of Human Control and Privacy}}}: For users to trust an AI system, user should be provided explicability of models and data usage, and how their data is collected and extensively used in the decision-making process \cite{ellison2011negotiating}. However, without the users' knowledge, majority of the data used to train AI algorithms and application systems is scraped from the internet and sold from third-party users \cite{buccinca2021trust, wu2023unveiling}.  \\
{\color{red}{\textbf{CASE STUDY}}}: In healthcare, AI systems are data driven and require leverage of confidential, discretional and sensitive patient data (e.g., age, sexual and gender orientation), including sensitive medical information (e.g., genetic diseases). Consequently, this can cause potential ethical and privacy concerns as data may no longer be confidential, and may be accessed unwilling.

\subsection{Proliferation of Malicious Intent and Manipulation}
\noindent{\textit{\textbf{Ethical Concerns}}}: As overreliance and dependency on automation increase due to the ubiquitous exposure, adoption, implementation and usage of AI systems is becoming normalized in society, ethical concerns will also continue to increase as a result of overestimating the potentialities of AI generated content \cite{overreliance}. This can lead to the proliferation of harmful biases, prejudice, discrimination, etc., in both an online (data poisoning and backdoor attacks \cite{chen2017targeted, tolpegin2020data, tran2018spectral}) and offline (stereotypes and representations harms \cite{10.5555/3157382.3157584, 10.1145/3442442.3452325, dacon2022detecting, tatman17_interspeech}) settings. \\

\noindent{\textit{\textbf{Physical and Digital Harm}}}: As previously mentioned, the proliferation of ethical issues presents a existential risk since social platforms allows a myriad of harmful practices, particularly, from individuals or social groups who may oppose, criticize, or possess contradictory feelings, beliefs, or motivations towards certain social groups or communities (e.g., LGBTQIA2S+ and BIPOC (Black, Indigenous, and People of Color) \cite{dacon2022detecting, yin2021towards, waseem-hovy-2016-hateful}. Physical and digital harm is well studied in the AI research community, however, due to the anonymity of social platforms various types of physical (violence, harassment and abuse) \cite{nozza-etal-2021-honest} and cyberattacks are amplified \cite{zou2023universal, wu2023unveiling}. For example, in online scenarios in the form of non-consensual (or fake) imagery or audio, and offline setting in the form of verbal assaults (i.e., death threats, obscenity, insults, and identity-based attacks), harassment and abuse.\\

\noindent{\textit{\textbf{Information Disorder}}}: Due to the inability of AI systems to act on generated content, with inadvertent overuse of social platforms and ubiquitous exposure to media content ``\textit{humans may put a higher degree of trust in AI generated content, especially when outputs appear authoritative or when people are in time-sensitive situations}'' \cite{solaiman2023evaluating, wardle2017information}. Therefore, consumers of media content can be victims of information disorder (misinformation, disinformation and malinformation). These terms may be used interchangeably, yet they have subtle differences.

\begin{itemize}
    \item[--] Misinformation: The unconscious distribution of inaccurate information without the intent to mislead e.g., statistics, dates, language translations, etc.
    \item[--] Disinformation: The conscious distribution of biased information with the malicious intent of deliberately misleading e.g., fake news, political propaganda, rumors, conspiracy theories, etc. 
    \item[--] Malinformation: The conscious distribution of information with the malicious intent of deliberate and strategic manipulation with a conjoint objective of inflicting harm or causing advantage using sensitive, and confidential information, and releasing consensual information non-consensually e.g., email phishing, revenge porn, etc.
\end{itemize}

At a global scale, the coupled relationship of the monopolization of consumer data and information pollution have become a complex social issues as a myriad of AI application systems (i.e., social media platforms, web forums and generative AI systems) expedite and simplify amplifying content types and techniques \cite{wardle2017information}.

\section{Conclusion}\label{sec:conclusion}

In this work, we present a comprehensive overview of pressing concerns of Trustworthy AI from a socioethical impact assessment perspective to explicate its harmful societal effects on the environment and human society. We believe that it is crucial to disclose ecological and sociological concerns of Trustworthy AI research in full, hence, we focus on sociological and ecological concerns surrounding widely adopted AI system's trustworthiness to further deepen the understanding of the adverse societal impacts of AI. We provide numerous examples of real-world impacts and highlight several potential issues that may undermine our trust in AI systems.  We --- users (or consumers), auditors, legislators and policymakers, and researchers and developers of digital tools need become aware the energy to develop and consume large-scale AI application systems as datasets and models become more complex, thus, contributing to a substantial water and carbon footprint. This article contributes towards this objective. In this work, we offered a brief, high-level view of societal impacts of widely adopted and deployed trustworthy AI systems by adopting a proactive approach to accentuate that AI systems require multi-disciplinary governance and convergence, and cross-sectoral cooperation throughout its lifecycle to consider its induced sociological and ecological effects in human society and the environment.
Finally, we hope this article will inspire developers of AI/ML/NLP algorithms, models and application systems to disclose the adverse environmental and social impacts of their models, so that such information can be monitored via a possess a human-in-the-loop, human-on-the-loop, or human-in-command paradigm alongside model accuracy, emission and consumption metrics to truly enable humanity-centered Trustworthy AI.

\bibliographystyle{ACM-Reference-Format}
\bibliography{main}


\end{document}